\documentclass[12pt]{article}
\usepackage{psfig}

\def\lesssim{\mathrel{\mathpalette\vereq<}}
\def\gtrsim{\mathrel{\mathpalette\vereq>}}
\makeatletter
\def\vereq#1#2{\lower3pt\vbox{\baselineskip1.5pt \lineskip1.5pt
\ialign{$\m@th#1\hfill##\hfil$\crcr#2\crcr\sim\crcr}}}
\makeatother

\begin{document}

\begin{titlepage}
\begin{center}
\today     \hfill    LBNL-41199 \\
~{} \hfill UCB-PTH-97/69  \\
~{} \hfill hep-ph/9712515\\

\vskip .1in

{\large \bf Sneutrino Cold Dark Matter\\
With Lepton-Number Violation}%
\footnote{This work was supported in part by the U.S. 
Department of Energy under Contracts DE-AC03-76SF00098, in part by the 
National Science Foundation under grant PHY-95-14797.  HM was also 
supported by Alfred P. Sloan Foundation.}

\vskip 0.3in

Lawrence J. Hall,$^{1,2}$ Takeo Moroi$^{1}$ and Hitoshi Murayama$^{1,2}$

\vskip 0.05in

{\em $^{1}$Theoretical Physics Group\\
     Ernest Orlando Lawrence Berkeley National Laboratory\\
     University of California, Berkeley, California 94720}

\vskip 0.05in

and

\vskip 0.05in

{\em $^{2}$Department of Physics\\
     University of California, Berkeley, California 94720}

\end{center}

\vskip .1in

\begin{abstract}
The tau sneutrino is proposed as a candidate for galactic halo dark matter, 
and as the cold dark matter (CDM) component of the universe.
A lepton-number-violating sneutrino mass, $\tilde{\nu} \tilde{\nu}$,
splits the tau sneutrino into two mass eigenstates: $\tilde{\nu}
\rightarrow \tilde{\nu}_\pm$. The absence of a $Z \tilde{\nu}_-
\tilde{\nu}_- $ coupling implies that the lighter mass eignestate, 
$\tilde{\nu}_-$, does not annihilate via the $s$-channel $Z$-exchange
to a low cosmological 
abundance, and furthermore, halo sneutrinos do not scatter excessively
in Ge detectors. For the majority of the relevant parameter space, the
event rate in Ge detectors is $\geq 10^{-2}$ events/kg/day. The lepton number 
violation required for sneutrino CDM implies that the
tau neutrino mass is $m_{\nu_\tau} \gtrsim 5$ MeV, large enough to be
excluded by $B$ factory
experiments. Events of the form $l^+ l^-
\!\!\not\!\! E $ or 
$jj \!\!\not\!\! E $, with low $m_{ll}$ or $m_{jj}$, 
may be observed at LEP2. A seesaw
mechanism is investigated as the origin for the lepton number
violation, and several other cosmological and particle physics
consequences of sneutrino CDM are discussed. 
\end{abstract}

\end{titlepage}

\newpage

\noindent {\bf 1. Introduction.}
It has been known for decades that most of the mass in the Universe is
dark, {\it i.e.}\/ not seen by optical methods~\cite{dark}.
This is deduced from studying the motion of visible objects, which is
governed by the size of the gravitational force acting on them.
Rotational curves of spiral galaxies and
motions of galaxies in clusters are good examples, both of which
indicate that most sources of gravity are not seen.

Recently, MACHOs (MAssive Compact Halo Object) were seen within the 
halo of the Milky Way galaxy by means of gravitational microlensing 
\cite{MACHO}.  However, the determination of the MACHO mass fraction 
in the halo is still quite uncertain: anywhere between 10\% to 100\%.  
Moreover, the scenario of 100\% MACHO fraction faces various 
astrophysical and cosmological difficulties (see, {\it e.g.}\/,
\cite{noMACHO}).  Therefore it is quite 
possible that the MACHOs account for the missing dark baryons, as 
required by Big Bang Nucleosynthesis, but are not the dominant component of 
the galactic halo.

From the point of view of galaxy formation theories, and the
small density fluctuations observed by COBE,
the most promising candidate for the invisible source of 
gravity is Cold Dark Matter (CDM) \cite{wss}. 
Although the standard CDM model, with 
scale-invariant primordial density fluctuations, is not favored by the 
COBE data and the observed large scale structures, the small 
discrepancy can be accounted for by introducing a small Hot Dark Matter 
component~\cite{aph9707285}, by ``tilting'' the primordial density 
fluctuation spectrum~\cite{tilt}, or by introducing particles (such as 
$\nu_{\tau}$) whose decay changes the time of radiation-matter 
equality~\cite{cdm+mnu}.  In all these scenarios, CDM is the 
dominant component of the galactic halo.

There is no CDM candidate in the standard model.  On the other hand, 
theories of weak-scale supersymmetry are strongly motivated: they 
allow a symmetry description of the weak scale, they incorporate the 
economical description of flavor symmetry 
breaking by Yukawa couplings, and they successfully predict the weak
mixing angle, at the 
percent level, from gauge coupling unification.  Finally, the lightest 
superpartner (LSP) is a candidate for CDM.\footnote{We assume that its 
stability is guaranteed by $R$-parity.}

There are two obvious choices for a neutral LSP candidate for CDM:
neutralinos and sneutrinos. The neutralino candidate, especially the
case of the superpartner of the $U(1)_Y$ hypercharge gauge boson, the
bino $\tilde{B}$, has received extensive discussion
\cite{PRep267-195}.  For certain choice of superpartner masses, the
cosmological $\tilde{B}$ energy density can have the correct order of
magnitude to be the dark matter. Its interactions in semiconductor
detectors are sufficiently weak that it is an experimental challenge
to directly detect this form of CDM; its detection rate can be as low
as $10^{-4}$~events/kg/day.  Higgsino-like \cite{DNRY} and
mixed gaugino-Higgsino LSPs are also possible neutralino candidates
for CDM.

Sneutrinos annihilate rapidly in the early universe via $s$-channel 
$Z$ and $t$-channel neutralino and chargino exchange.  To reduce these 
annihilations and obtain a cosmologically significant 
$\Omega_{\tilde{\nu}}$, it was proposed that the sneutrinos should be 
light, $m_{\tilde{\nu}} \approx 2$~GeV \cite{i}.  Such light LSP 
sneutrinos could be obtained in minimal supergravity models, although 
from todays perspective such small scalar masses appear somewhat 
fine-tuned.  This light $\tilde{\nu}$ CDM has been 
excluded from measurements of the $Z$ width.
In supersymmetric models, a LSP sneutrino is 
expected to have a mass in the range of, say, 30---200~GeV from naturalness 
arguments. However, in this 
case the cosmological annihilation is large, leading to a low 
abundance.  The annihilation in the early universe can be reduced by 
taking the sneutrino heavier, 550---2300~GeV for $0.1 \lesssim 
\Omega_{\tilde{\nu}} \lesssim 1$.  Such a heavy sneutrino, already 
disfavored on theoretical grounds, is firmly excluded by the nuclear 
recoil direct detection searches: the $t$-channel $Z$ exchange gives a 
cross section four times larger than the case of a Dirac neutrino, 
excluding all $m_{\tilde{\nu}}$ up to 17 TeV if the sneutrino is the 
dominant component of the halo \cite{FOS}.  Sneutrino CDM is
apparently firmly excluded.

This negative conclusion on sneutrino CDM is based on the implicit
assumption of lepton number conservation, which implies three mass
eigenstates of sneutrino, each described by a complex field.
It is well known that the phenomenology of neutrinos is greatly
changed by the addition of lepton number violation, and the same is
true for sneutrinos --- each complex field now represents two
particles with different masses: $\tilde{\nu}_{\pm}$. In the minimal
standard model, without right-handed neutrinos, gauge invariance and
renormalizability ensure that the lepton numbers are exact symmetries.
However, the standard model is surely just a low energy effective
theory, and physics from high mass scales $M$ can induce lepton number
violation via the operator $llhh/M$, giving Majorana neutrino masses,
where $l$ and $h$ are lepton and
Higgs doublets. In supersymmetric extensions of the standard model,
with minimal field content and $R$-parity conservation, lepton number
violation can occur at dimension four by the operator $\tilde{l}
\tilde{l} hh$, which breaks supersymmetry explicitly, and gives a mass 
splitting between $\tilde{\nu}_{\pm}$.

\noindent{\bf 2. Phenomenology of $\tilde{\nu}$ CDM.}
The purpose of this letter is to present a phenomenological analysis 
on the viability of the sneutrino CDM with lepton-number violation.  
The sneutrinos carry the same lepton numbers as their supersymmetric
partners (neutrinos), and are distinguished from their anti-particles,
anti-sneutrinos.  They have soft supersymmetry-breaking masses which are
expected to be in the range of 30---200~GeV$/c^{2}$.  In the
presence of lepton number violation, sneutrinos can mix with
anti-sneutrinos because there are no other quantum numbers which forbid
the mixing \cite{Hirsch,GH}.  Without loss of
generality, the mass-squared matrix for a single generation of
sneutrinos can be parameterized by two real parameters,
$m_{\tilde{\nu}}^2$ and $\Delta m^2$:
\begin{equation}
        {\cal L}_{\rm mass} = \frac{1}{2}
                (\tilde{\nu}^{*}, \tilde{\nu}) \left( \begin{array}{cc}
                m_{\tilde{\nu}}^{2} & \Delta m^{2}/2 \\
                \Delta m^{2}/2 & m_{\tilde{\nu}}^{2} \end{array}
                \right)
                \left( \begin{array}{c} \tilde{\nu} \\ \tilde{\nu}^{*}
                \end{array} \right) ,
        \label{eq:mass-matrix} 
\end{equation}
where the positive mixing parameter $\Delta m^2$ is a consequence of the 
operator $\tilde{l}\tilde{l} h h $ mentioned earlier.
We later identify these as the tau sneutrinos.
We assume that $m_{\tilde{\nu}}^2$ is sufficiently positive that the physical 
mass eigenstates sneutrinos are
$\tilde{\nu}_{+} = (\tilde{\nu} + \tilde{\nu}^{*})/\sqrt{2}$
and $\tilde{\nu}_{-} = i(\tilde{\nu} - \tilde{\nu}^{*})/\sqrt{2}$,
with eigenvalues $m^{2}_{\tilde{\nu}_\pm} = m_{\tilde{\nu}}^{2} \pm
\Delta m^{2}/2$. The mass difference between $\tilde{\nu}_{-}$
and $\tilde{\nu}_{+}$ is
\begin{equation}
        \Delta m \equiv m_{\tilde{\nu}_+}-m_{\tilde{\nu}_-}        
\simeq \frac{\Delta m^2}{2m_{\tilde{\nu}}}
\end{equation}
for $\Delta m^2 \ll m_{\tilde{\nu}}^2$.

For our purpose, the most important property of the mass eigenstates
$\tilde{\nu}_{\pm}$ is that there is no diagonal coupling to the
$Z$-boson; its coupling is always off-diagonal, {\it i.e.}\/,
$Z$-$\tilde{\nu}_{+}$-$\tilde{\nu}_{-}$.  This result is a simple consequence
of Bose symmetry, and has a crucial impact on both the cosmological 
sneutrino abundance and on the signal for direct detection of halo sneutrinos.
With lepton number conservation, a large contribution to 
cosmological sneutrino annihilation comes from the $s$-channel exchange of a
virtual $Z$ boson, $\tilde{\nu} \tilde{\nu}^{*} \rightarrow Z^{*}
\rightarrow f\bar{f}$, where $f$ is any of the Standard Model quarks
and leptons with kinematically allowed masses. Although
the annihilation process is  $P$-wave, the large number of allowed
final states and fixed $m_Z$ makes this process important: $\sum_f
\sigma(\tilde{\nu} \tilde{\nu}^* \rightarrow f\bar{f}) = 0.0072 v_{rel}
m_{\tilde{\nu}}^2/(4 m_{\tilde{\nu}}^2 - m_Z^2)^2$, where $v_{rel}$ is
the relative velocity of the two sneutrinos.  With lepton-number
violation, however, this process is replaced by the {\it
  co}\/-annihilation $\tilde{\nu}_{+} \tilde{\nu}_{-} \rightarrow
f\bar{f}$.  Unless the mass splitting is too small $\Delta m\lesssim
5$~GeV (see below), the annihilation via the $s$-channel $Z$-exchange
can be suppressed effectively.  Moreover, the mass splitting between
the sneutrino and slepton in the same multiplet is given by the
$D$-term, $m_{\tilde{l}}^2 - m_{\tilde{\nu}}^2 =
(1-\sin^2\theta_W)m_Z^2 (-\cos2\beta)>0$, which is quite important for 
$m_{\tilde{\nu}_{-}}< m_{W}$ and a
moderately large ratio of Higgs vevs $\tan\beta \gtrsim 2$.  Then the
coannihilation with 
the charged slepton becomes unimportant.  Therefore, the dominant annihilation
process is via the $t$- and $u$-channel neutralino exchange, to which
we will return shortly.  If $m_{\tilde{\nu}} > m_{W}$, however,
other processes $\tilde{\nu} \tilde{\nu}^{*} \rightarrow W^{-} W^{+},
ZZ$ are possible via $t$-channel slepton or sneutrino exchange.  In
this case the mixing of the sneutrinos does not affect the
annihilation process significantly and hence the earlier analyses
\cite{FOS} apply.  Therefore, we focus on the range
$m_{\tilde{\nu}_{-}} < m_{W}$.  We also assume that
$\tilde{\nu}_{-}\tilde{\nu}_{-} \rightarrow h h$ is not kinematically
allowed. 

\begin{figure}[tbp]
   \begin{center}
     \leavevmode
     \psfig{file=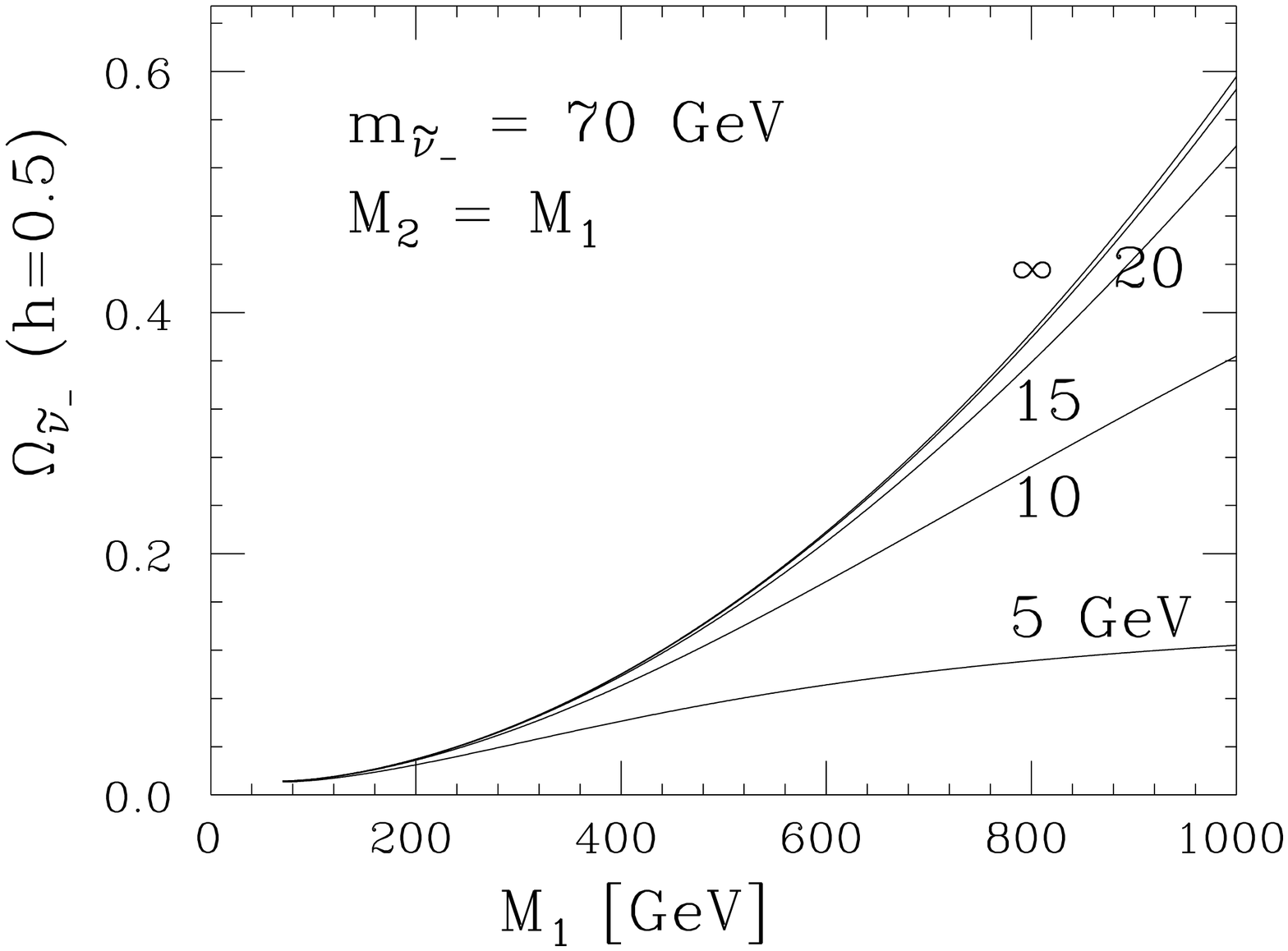,width=0.45\textwidth,angle=90}
     \psfig{file=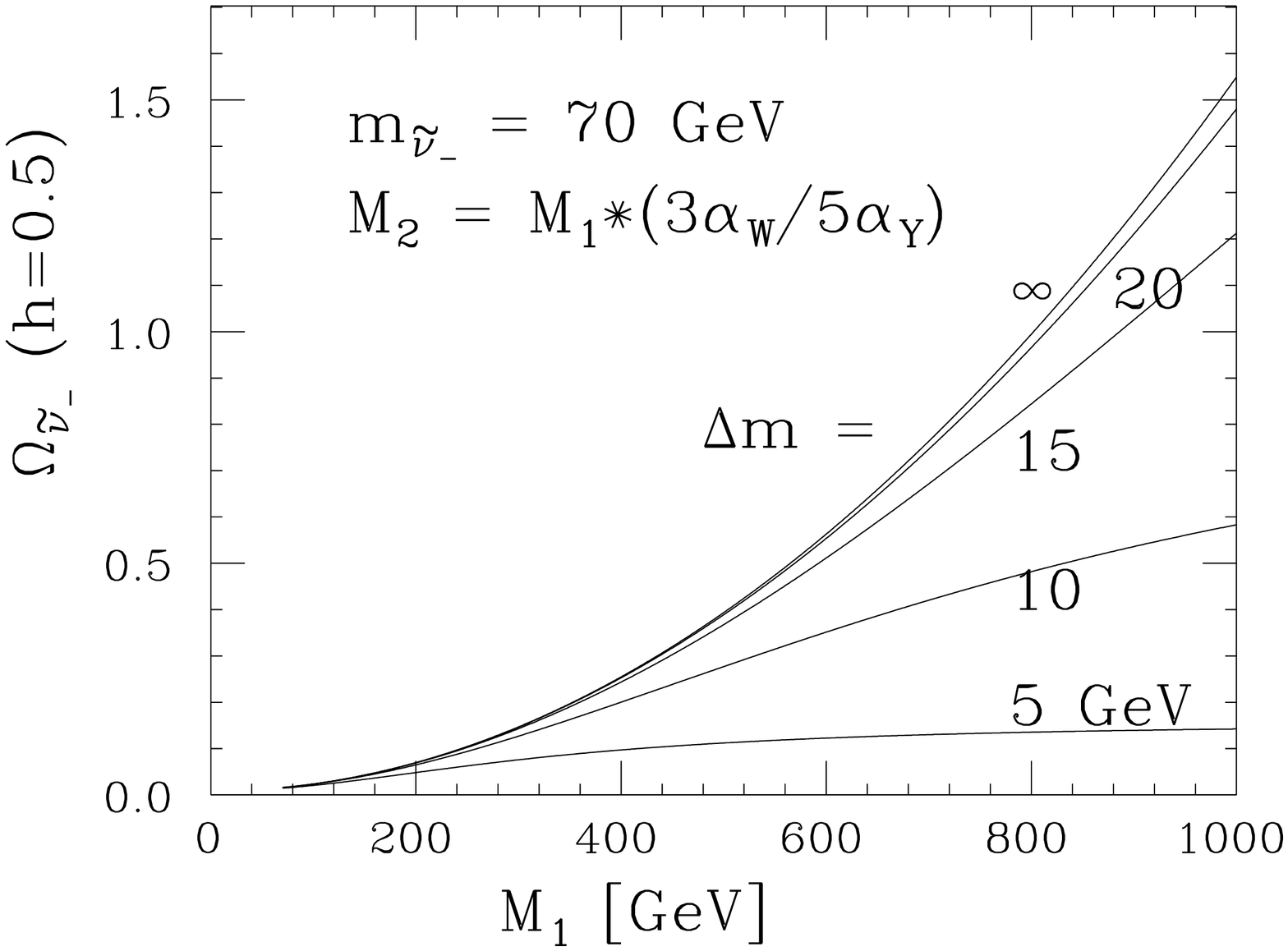,width=0.45\textwidth,angle=90}\\
     \leavevmode
     \psfig{file=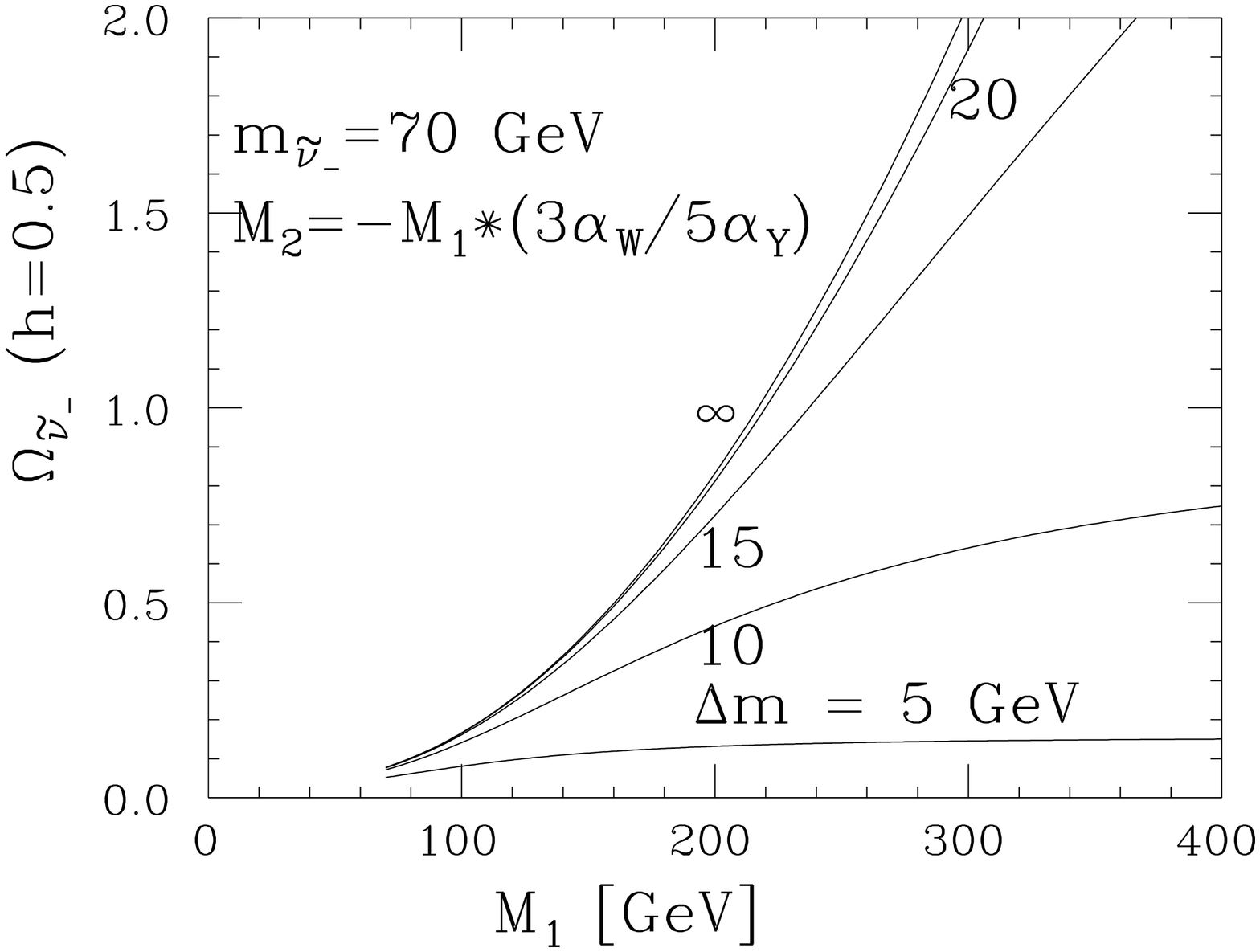,width=0.45\textwidth,angle=90}
     \psfig{file=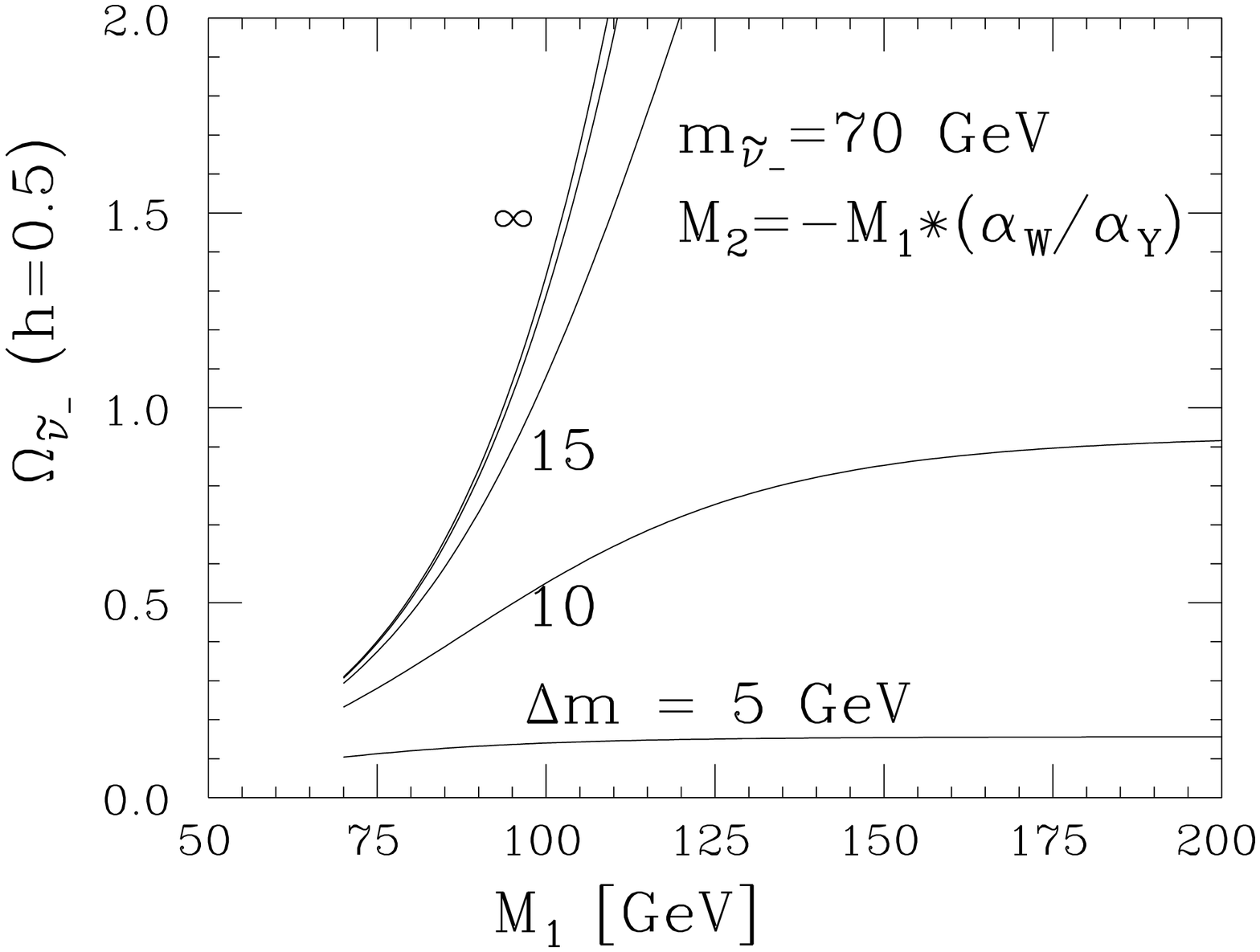,width=0.45\textwidth,angle=90}
     \label{fig:relic}
   \end{center}
  \caption[relic]{The present relic density of the sneutrino
    $\Omega_{\tilde{\nu}}$ with $H_0 = 50$~Mpc/km/s, with
    $m_{\tilde{\nu}_-} = 70$~GeV and $\Delta m = 5, 10, 15,
    20$~GeV and $\infty$.  The annihilation processes included
    are $t$- and 
    $u$-channel $\tilde{B}$, $\tilde{W}^3$ exchange and coannhilation
    with $\tilde{\nu}_+$ via $s$-channel $Z$-exchange.  Four plots
    assume different ratios of $M_1$ and $M_2$ as quoted.  Note the
    different scales in the plots.}
\end{figure}

The calculation of the cosmic abundance is
standard~\cite{Kolb&Turner}.  In Fig.~1, we show the values of
$\Omega_{\tilde{\nu}}$ for $H_0 = 50$~Mpc/km/s ($h_0 = 0.5$) as
functions of $M_1$ with various values of $\Delta m$.  We considered
the most important annihilation processes as discussed above: $t$- and
$u$-channel exchange of the bino $\tilde{B}$ and the neutral wino
$\tilde{W}^3$, with the cross section
\begin{equation}
  v_{rel} \sigma(\tilde{\nu}_- \tilde{\nu}_- \rightarrow \nu\nu,
  \bar{\nu}\bar{\nu})
  = \frac{\pi}{4} \left( \frac{\alpha_Y M_1}{m_{\tilde{\nu}_-}^2+M_1^2}
    + \frac{\alpha_W M_2}{m_{\tilde{\nu}_-}^2+M_2^2} \right)^2 ,
\end{equation}
and also the coannihilation effect with $s$-channel $Z$-exchange
suppressed by the Boltzman factor $e^{-\Delta m/T}$.  The temperature
is taken at the annihilation freezeout: $T \simeq m/25$.  Note that the thermal
average over the initial state should include the statistical factor
$1/2!$ to avoid double counting of states in the Boltzmann equation.
The cross section depends sensitively on the relative ratio (and sign)
of $M_1$ and $M_2$.  The $SU(5)$ grand-unified theory predicts $M_2 =
M_1\times (3\alpha_W/5\alpha_Y)$.  We vary the ratio freely for the
purpose of our phenomenological analysis.  There is also a
contribution from the $s$-channel Higgs boson exchange into $b\bar{b}$
or $\tau^+ \tau^-$, but we have checked that it is always much smaller
than the neutralino exchange for the range shown in the plot.  Recall
that $\Omega \sim 0.03$--$0.4$ is needed for halo dark matter, while
measurements at larger scales suggest somewhat larger range.
Inflation predicts $\Omega = 1$.  With the grand-unified gaugino
mass relation, the range required for halo dark matter can be obtained with
$M_1 \gtrsim 200$~GeV and $\Delta m \gtrsim 5$~GeV.  With more general
gaugino mass parameters, even the value preferred by inflation can be
easily obtained. 
Lepton-number violation allows the sneutrino to become a viable CDM candidate.

Next we consider the detection of galactic halo sneutrinos in Ge detectors.  
The scattering of $\tilde{\nu}_{-}$ cannot produce $\tilde{\nu}_{+}$ 
due to simple kinematics if $\Delta m > \beta_h^2 m_{\tilde{\nu}_-} m_A
/2(m_{\tilde{\nu}_-} + m_A) = 20$~keV
for $m_{\tilde{\nu}_-}= m_W$, $m_A = 
72$~GeV for Ge, and $\beta_h = 10^{-3}$ for virialized halo 
particles on average.  Therefore, there is no $Z$-exchange process beween the 
sneutrino and the nucleus, and hence the bound from the direct 
detection experiment described earlier does not apply.\footnote{The absence 
of the $Z \tilde{\nu}_- \tilde{\nu}_-$ coupling also implies that fewer halo 
sneutrinos are captured by the sun. We find that present limits on high 
energy neutrinos from the sun do not place a constraint on our scheme.} 
The dominant 
contribution to the scattering comes from the lightest Higgs boson 
exchange.  We assume that the heavy Higgs boson, whose exchange 
may enhance the cross section, is sufficiently heavy such that the lightest
Higgs boson has the same coupling as the Standard Model Higgs boson.
The coupling of the Higgs boson  
and the sneutrino comes from the $D$-term potential in the 
supersymmetric Lagrangian as well as from the SUSY breaking operator 
$\tilde{l}\tilde{l}hh$,  which is also the origin of the sneutrino mass 
splitting $\Delta m^2$. For $\tan\beta>1$, these two contributions always 
interfere constructively, 
so that the scattering cross section can be estimated as
\begin{equation}
        \sigma = \frac{1}{81\pi(m_{\tilde{\nu}_-} + m_{A})^2}\left(
   \frac{m_{A}^2(\Delta m^2 - m_{Z}^2 \cos 2\beta)}{v^2 m_{h}^2}
   \right)^{2} ,
\end{equation}
with $v=250$~GeV.  Recall that the lightest Higgs boson
in the Minimal Supersymmetric Standard Model has to be lighter than
130~GeV$/c^{2}$ and must be in a comparable range even in
non-minimal extensions, if perturbativity up to the Planck scale is
assumed \cite{MO}. 

\begin{figure}[tbp]
   \begin{center}
     \leavevmode
     \psfig{file=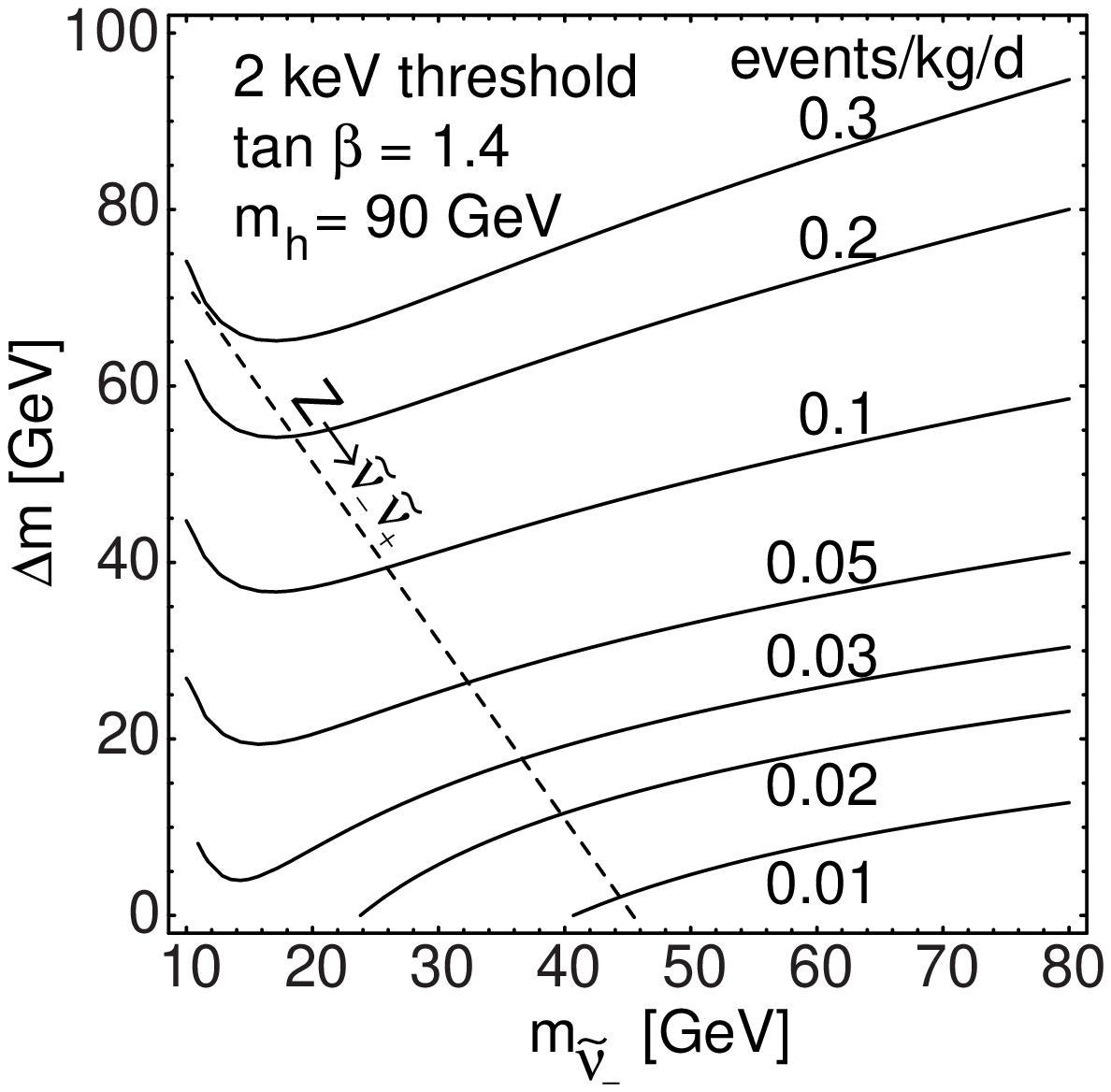,width=0.45\textwidth}
     \psfig{file=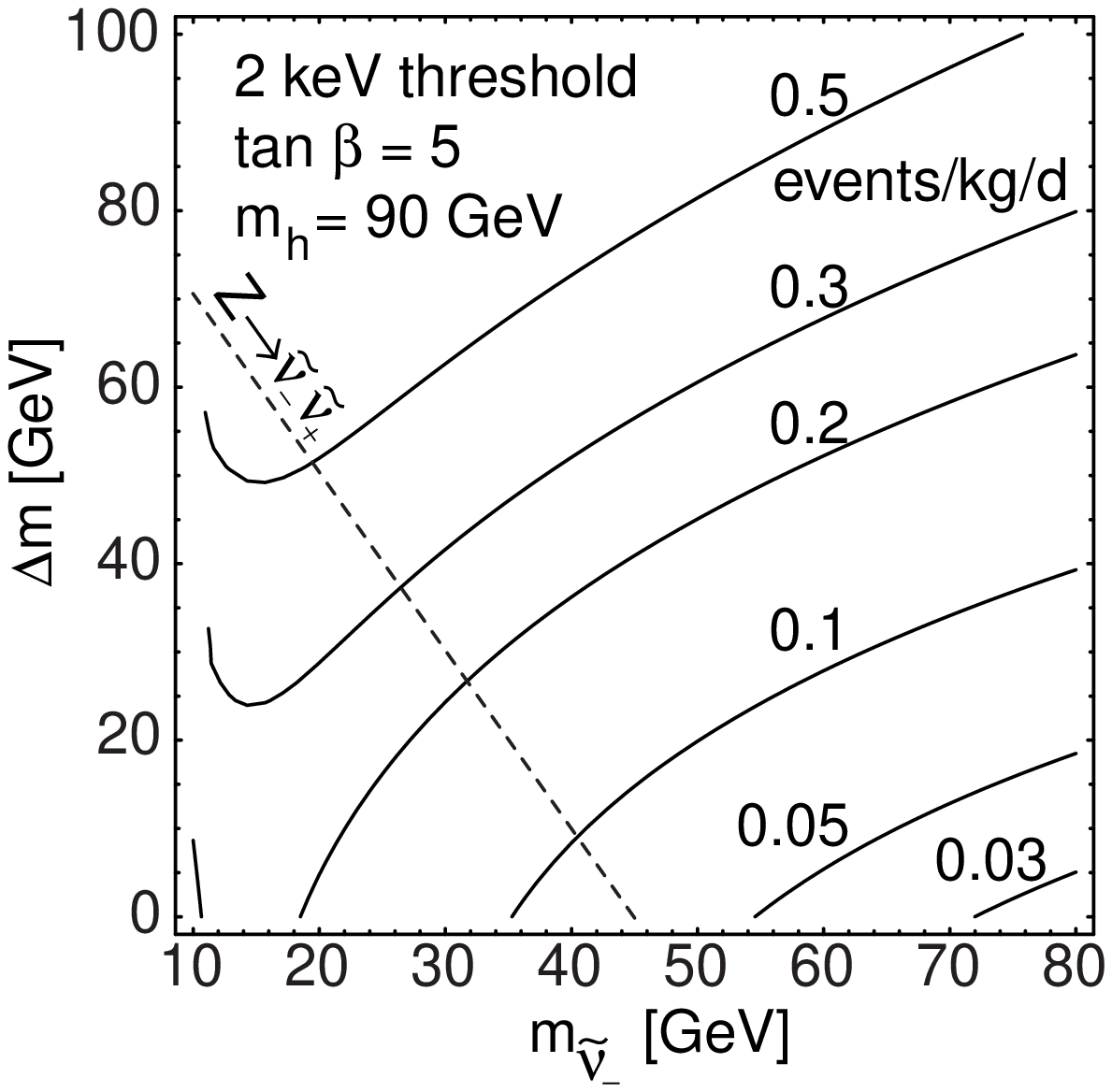,width=0.45\textwidth}
     \label{fig:det}
   \end{center}
  \caption[relic]{The detection rate of sneutrino CDM in Ge
    detectors, in units of events/kg/day, for a halo density 
    $\rho = 0.3$~GeV$/\mbox{cm}^{3}$, and $m_h=90$~GeV.  The threshold
    energy of the detector is assumed to be 2~keV.  We take
    $\tan\beta=1.4$ and 5.  The dotted line shows the contour for
    $m_{\tilde{\nu}_-} +m_{\tilde{\nu}_+}=m_Z$, and hence the region below
    it is excluded by LEP1.}
\end{figure}

We show the counting rate of sneutrino CDM with Ge detectors in 
Fig.~2.  Here, we assume the local halo density $\rho = 
0.3$~GeV$/\mbox{cm}^{3}$, lightest Higgs mass 
$m_h=90$~GeV, and the isothermal distribution of halo particles
with the average velocity $\beta_h = 10^{-3}$.  The lowest value of
$\tan\beta$ which keeps the top 
Yukawa coupling perturbative up to the GUT-scale is 1.4 and we used
this value as the case with lowest possible detection rate.  Another
case shown is $\tan\beta=5$.  For larger $\tan\beta$ the detection
rate is somewhat larger than the latter case but not much. To obtain a
large enough relic  
density of the sneutrino, $\Delta m$ has to be larger than 
(5--10)~GeV; otherwise coannihilation effect reduces the sneutrino 
abundance irrespective of the gaugino mass.  In this region, the 
detection rate can be typically $10^{-2}/\mbox{kg}/\mbox{day}$ or 
larger, which is within the reach of future detection of the CDM at Ge 
detector.\footnote{We also estimated the nuclear form factor
  suppression using the formula in \cite{DN}.  The suppression factor
  is always less than a factor of two for this light range of CDM
  mass.}  
If $m_h$ is increased to its maximum value of 135 GeV, these rates are
decreased by about a factor of four. Nevertheless, direct detection
searches are able to probe a larger fraction of the relevant parameter
space for sneutrino CDM than for neutralino CDM.

It is an important question what part of the $(m_{\tilde{\nu}_{-}}, 
\Delta m)$ parameter space is allowed by current collider 
experiments.  
For $\Delta m$ large enough to give a significant cosmological abundance, 
any $\tilde{\nu}_{+}$ produced at colliders will decay into 
$\tilde{\nu}_- j j$ or $\tilde{\nu}_- l^+ l^-$ inside the detector.  
The signature at LEP1 and LEP2 results from the pair 
production $e^+ e^- \rightarrow \tilde{\nu}_- \tilde{\nu}_+$ via 
$s$-channel $Z$-exchange, with a subsequent decay $\tilde{\nu}_+ 
\rightarrow  
\tilde{\nu}_- j j$ or $\tilde{\nu}_- l^+ l^-$.  This is similar to the 
signature of the higgsino-like neutralino $\tilde{\chi}_1^0 
\tilde{\chi}_2^0$ production and the subsequent decay of 
$\tilde{\chi}_2^0$.  The LEP1 constraint is basically  
$m_{\tilde{\nu}_{-}}+m_{\tilde{\nu}_{+}} = 2 m_{\tilde{\nu}_{-}}+\Delta 
m < m_{Z}$ \cite{L3}.  
This limit is shown in Fig.~2.
 
\begin{figure}
  \begin{center}
     \leavevmode
     \psfig{file=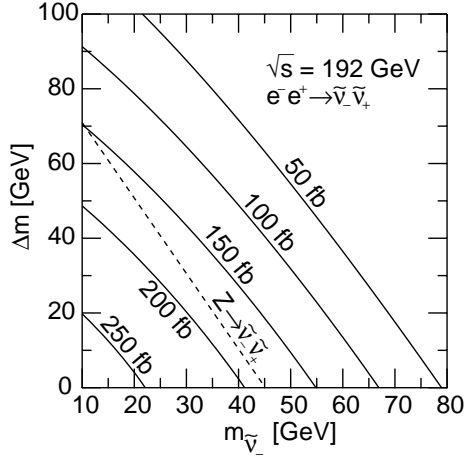,width=0.45\textwidth}
   \end{center} \caption[cross]{The production cross section for $e^{+}
e^{-} \rightarrow \tilde{\nu}_{+} \tilde{\nu}_{-}$ at
$\sqrt{s}=192$~GeV (solid lines). The dotted line shows the contour for
$m_{\tilde{\nu}_-} +m_{\tilde{\nu}_+}=m_Z$, and hence the region below the
dotted line is excluded by LEP1.}
\label{fig:cross}
\end{figure}

At $\sqrt{s}=172$~GeV, we find the cross section 
$\sigma(e^{+}e^{-}\rightarrow\tilde{\nu}_{+}\tilde{\nu}_{-})$ 
to be smaller than $\sim 300$~fb, 
while the current upper bound on the neutralino 
production cross section is about 800~fb or 
larger~\cite{opal_neutralino}. Hence the sneutrino LSP is not 
constrained by this bound.  However, in the near future,
with the full luminosity of LEP2, the cross section will be 
constrained to be below 100--200~fb~\cite{GMRR}. 
In  Fig.~\ref{fig:cross} the cross section  
$\sigma(e^{+}e^{-}\rightarrow\tilde{\nu}_{+}\tilde{\nu}_{-})$ is shown
for $\sqrt{s}=192$~GeV: LEP2 may probe a significant portion of the
interesting parameter space.

So far we have phenomenologically parametrized the sneutrino mass
matrix by varying $m_{\tilde{\nu}}^{2}$ and $\Delta m^{2}$ freely.
One cannot, however, make $\Delta m^{2}$ arbitrarily large because the
lepton number violation in the sneutrino mass matrix induces a
Majorana mass for its partner neutrino from one-loop diagrams.  The
authors of Ref.~\cite{GH} analyzed this question and
found\footnote{This result assumes no accidental cancellations between
  tree level and various one-loop contributions to the neutrino mass.} 
\begin{equation}
        m_{\nu} \gtrsim \frac{\Delta m}{2\times 10^{3}}.
\end{equation}
 Given the value of the $\Delta m^{2}$ necessary to keep a large
enough cosmic abdundance of sneutrino, we conclude that the sneutrino
CDM must be the tau-sneutrino.  The current limit on the tau neutrino
mass is $m_{\nu_\tau} <
18.2$~MeV$/c^{2}$ \cite{nutaumass}, so there is room for
$\Delta m^{2}$ in the cosmologically interesting range. An important
consequence of sneutrino CDM is that the $\nu_{\tau}$
mass is in the region of 10~MeV.  The asymmetric $B$-factory
experiments at SLAC and KEK, BABAR and BELLE, will be able to exclude
the finite $\nu_{\tau}$ mass down to 2~MeV$/c^{2}$ 
range \cite{Seiden}.
It will be particularly interesting if both the direct search
experiments for CDM see a signal and BABAR or BELLE measure a
finite $\nu_{\tau}$ mass.  It would be possible to study the
consistency of the two results to determine the underlying parameter set.

\noindent{\bf 3. A model with right-handed neutrinos.}
What underlying theory of lepton number violation could lead to
$\Delta m \approx 10$ GeV? Since $\tilde{\nu}_-$ must be stable, we
seek an $R$-parity conserving origin for the $\Delta L =2$ operator
$O_1 = [llhhz]_F/\Lambda$ where $h$ is the up-type Higgs doublet and
$z$ is a dimensionless spurion 
field $z=A \theta^2$.\footnote{Here and below, $[\ldots]_F$ refers to the
  $F$-component of the chiral superfield in square brackets.} This
operator gives $\Delta m = 
(A/m_{\tilde{\nu}}) (v^2 \sin^2\beta/\Lambda)$. 
In general one expects $O_1$ to be
accompanied by $O_2 = [llhh]_F/\Lambda$, which leads to $m_\nu = v^2
\sin^2\beta/\Lambda$ and 
the relation $\Delta m = (A/m_{\tilde{\nu}}) m_\nu$.  In theories with
supersymmetry broken in a hidden sector of supergravity, one finds
$(A/m_{\tilde{\nu}}) \approx 1$, giving $\Delta m \approx m_\nu \le
18.2$~MeV$/c^{2}$, which is a factor $10^3$ too small.

Suppose that operators $O_{1,2}$ arise on
integrating out a heavy right-handed neutrino, $\nu_R$, which has the
interactions $[\frac{1}{2}\lambda S \nu_R \nu_R + h_\nu l \nu_R
h]_F$. Heavy particles may be coupled to large supersymmetry breaking
without upsetting the gauge hierarchy, so that we consider $\langle
S\rangle = V + \theta^2 F$, a mass for the right-handed neutrino is
generated $M=\lambda V$.  The effective operators $O_1$, $O_2$ are
obtained upon integrating out the right-handed neutrino, with $\Lambda=
2 \lambda V / h_\nu^2$, $A=F/V$ and hence the relation
\begin{equation}
\Delta m = {F \over V m_{\tilde{\nu}}} m_\nu.
\end{equation}
Hence, if $\nu_R$ is coupled to a field $S$ which has $F/V \approx 10^3 
m_{\tilde{\nu}}$, then $\Delta m$ is sufficient to allow
$\tilde{\nu}_-$ to be CDM. Since $\nu_R$ is vector-like with respect
to the standard model gauge group, it is not surprising that it is
coupled to larger symmetry breakings than the light matter --- this is
the motivation for the seesaw mechanism itself --- however, we have no
convincing argument for the magnitude of $F/V$.

There is an important constraint on the scale $M$.
The sum of the tree-level neutrino mass and the one-loop induced
term should not be larger than the experimental limit of
18.2~MeV/$c^2$, and, barring a possible cancellation, we require
$m_\nu^{\rm tree} = (h_\nu v\sin\beta)^2/2M \lesssim 20$~MeV which bounds $M$
from below.  On the other hand, a sufficient $\Omega_{\tilde{\nu}}$
requires  $\Delta m^2 = 2 A
m_\nu^{\rm tree} \gtrsim 500~\mbox{GeV}^2$.
This requires $A \gtrsim .0075 \cdot M/h_\nu^2$.  Finally, the large
supersymmetry breaking $A$ in the right-handed neutrino generates
corrections to $m^2_{\tilde{l}}$ and $m^2_{\tilde{h}}$ via two-loop
diagrams.  This can be calculated using the method of Giudice and
Rattazzi to leading order in $A$ \cite{GR}, and we find
\begin{eqnarray}
  \delta m_l^2 &=& \frac{1}{(16\pi^2)^2} A^2 h_\nu^2 (4 h_\nu^2 + 3
  h_t^2 - 3 g^2 - g'^2), \\
  \delta m_h^2 &=& \frac{1}{(16\pi^2)^2} A^2 h_\nu^2 (4 h_\nu^2 - 3
  g^2 - g'^2),
\end{eqnarray}
where $h_t$ is the top quark Yukawa coupling, and $g$, $g'$ are
$SU(2)\times U(1)$ gauge coupling constants. On naturalness grounds, these
corrections should not be larger than about $(100~\mbox{GeV})^2$.
Combined with the lower bound on $A$ found above, we find an upper
bound on $M$; in fact for $\delta m_{h,l}^2 < (100 \mbox{GeV})^2$
we find that the allowed region is given by
$M = 100$--700~TeV and $h_\nu = 0.2$--0.7.\footnote{The constraints
from $\delta m_l^2$, $\delta m_h^2$ are somewhat subjective.   Also a
different model of lepton number violation (such as a weak-triplet
lepton exchange generating $O_1$, $O_2$) leads to very different
results for $\delta m_l^2$, $\delta m_h^2$ and hence the constraints
here are model-dependent.}
It is interesting that the scale $M$ is comparable to the one found in the 
simplest theories of gauge mediation, 
so that $S$ can be identified as the 
singlet field which gives gauge mediated supersymmetry
breaking \cite{DNS}. However, for $\tilde{\nu}_-$ to be the LSP, it is
necessary 
that the gravitino mass be larger than  $m_{\tilde{\nu}_-}$, which
requires the existence of a larger primordial supersymmetry breaking
in the theory $F_P \ge 10^{10} \mbox{GeV} \gg F$.  This happens when
the messenger U(1) gauge coupling is somewhat small.  The gravitino
heavier than the sneutrino is actually cosmologically favorable
because the gravitino LSP is rather problematic \cite{Andre}.

\noindent{\bf 4. Conclusions}
In the minimal supersymmetric standard model, 
the sneutrino is firmly excluded
as a CDM candidate. The $Z \tilde{\nu}^\dagger \tilde{\nu}$ coupling
leads to rapid cosmological annihilation, and therefore low values of 
$\Omega_{\tilde{\nu}}$, unless $m_{\tilde{\nu}}$ is very large, in
which case the same coupling leads to a large and excluded 
event rate in Ge detectors of halo CDM particles. 
In this letter we have shown that sneutrino CDM is allowed in
supersymmetric theories with lepton number violation.
A lepton number violating sneutrino
mass implies that each flavor of sneutrino has two distinct mass
states $\tilde{\nu}_\pm$. In this case there is only an off-diagonal
$Z$ coupling, $Z \tilde{\nu}_+ \tilde{\nu}_-$, so that if the mass
splitting of these two states is larger than about 5 GeV, and if the
lightest sneutrino has a mass in the range of about 40---80 GeV, 
$\Omega_{\tilde{\nu}}$ in the interesting range of 0.1 to 1 can
result. We have shown that the seesaw mechanism, which gives small
neutrino masses from integrating out heavy right-handed neutrinos, can
also lead to the required lepton number violation in the sneutrino
mass matrix.

There are three important, pre-LHC/LC tests for sneutrino CDM:
\begin{itemize}
\item Galactic halo sneutrinos will scatter in Ge detectors with an
event rate $\geq 10^{-2}$ events/kg/day, for most of the relevant
parameter range (see Fig.~2).
\item $m_{\nu_\tau} \gtrsim 5$ MeV, unless different contributions to the
neutrino mass are fined tuned to cancel.  This mass range of
$\nu_\tau$ can be excluded by the $B$-factory experiments.
\item  Events of the form $l^+ l^- \!\not\!\! E$ or $jj \!\not\!\! E$, 
with low $m_{ll}$ or
$m_{jj}$, may be
observed at LEP2, with $\sqrt{s} =192$ GeV (see Fig.~3). They result from
$\tilde{\nu}_+ \tilde{\nu}_-$ pair production, followed by
$\tilde{\nu}_+$ decay.
\end{itemize}

There are further important consequences of sneutrino CDM: 

(1) There are important new collider signatures of
supersymmetry. Squark and gluino production at hadron colliders leads
to events with substantial missing transverse energy, which is carried
away by the undetected $\tilde{\nu}_-$. However, a large fraction of
these events have $\tilde{\nu}_+$ in the decay chain, and when these
decay to $\tilde{\nu}_-$ they can produce lepton pairs with small
invariant mass. This decay, $\tilde{\nu}_+ \rightarrow \tilde{\nu}_- l^+
l^-$, becomes an important characteristic feature of many
supersymmetric signals. Also, the lightest Higgs boson may decay
dominantly to sneutrinos, $h \rightarrow \tilde{\nu}_+ \tilde{\nu}_+, 
\tilde{\nu}_-\tilde{\nu}_-$. It is 
possible that  only the invisible $\tilde{\nu}_-
\tilde{\nu}_-$ channel is kinematically allowed.

(2) The $\nu_\tau$, with 
its mass in the expected (5--20) MeV range, would overclose the 
universe if it is stable.  A visible decay, such as $\nu_\tau 
\rightarrow \nu_{e,\mu}\gamma$ or $\nu_{e,\mu} e^+ e^-$, and 
the invisible $3\nu$ mode, are 
disfavored for a variety of reasons. The $\nu_\tau$ should decay into
a massless boson $\nu_\tau \rightarrow \nu_{e,\mu} f$, with $f$ a
Majoron or familon, whose phenomenology was discussed recently in
detail \cite{FMMS}. 
The existence of such a massless boson is natural if the lepton
number is broken spontaneously at the mass scale of right-handed
neutrino \cite{CMP}.  It is 
interesting to note that a $\nu_\tau$ in this mass range, and 
with lifetime $10^{-2}~\mbox{sec} \lesssim \tau_\nu \lesssim 1$~sec, 
improves the situation with Big-Bang Nucleosynthesis \cite{BBN}.  

(3) For a critical universe with $\Omega_{\tilde{\nu}} = 1$, 
unification of the gaugino mass parameters is strongly disfavored.

(4) In the early universe, the large lepton number violation in the neutrino 
sector, together with high temperature $B+L$ sphaleron transitions, may 
wash out the cosmological baryon asymmetry \cite{FY}. 
Hence the baryon asymmetry should either be
generated at low temperatures, beneath the electroweak phase 
transition, or protected from sphaleron washout by condensates \cite{dmo} 
or by other exact symmetries.

(5) The sneutrinos must be coupled more strongly to supersymmetry
breaking than occurs in the simplest supergravity models.
Such mediation of supersymmetry breaking can readily occur via the gauge 
singlet right-handed neutrino.

\noindent{\bf Acknowledgements} We thank Savas Dimopoulos for
reminding us of the sphaleron washout of the baryon asymmetry.  HM
thanks Abe Seiden and Bernard Sadoulet for useful conversations. 
This work was supported in part by the U.S. 
Department of Energy under Contracts DE-AC03-76SF00098, in part by the 
National Science Foundation under grant PHY-95-14797.  HM was also 
supported by Alfred P. Sloan Foundation.

\end{document}